\documentstyle[preprint,prb,aps]{revtex}
\tighten
\begin{document}

\draft
\title{Anomalous Coherent Backscattering of Light from Opal
Photonic Crystals}
\author{J. Huang, N. Eradat, M. E. Raikh, and Z. V. Vardeny}
\address{Department of Physics, University of Utah, Salt Lake City, 
Utah 84112}
\author{A. A. Zakhidov and R. H. Baughman}
\address{Honeywell Int., Research and Technology Center, Morristown,
New Jersey 07962}
\maketitle

\begin{abstract}
We studied coherent backscattering (CBS) of light from opal photonic
crystals in air at different incident inclination angles, wavelengths
and along various $\left[hkl\right]$ directions inside the opals.
Similar to previously obtained CBS  cones from various random media,
we found that when Bragg condition with the incident light beam is 
not met then the CBS cones from opals show a triangular line shape in
excellent agreement with light diffusion theory. 
At Bragg condition, however,
we observed a dramatic broadening of the opal CBS cones that depends
on the incident angle and $\left[hkl\right]$ direction. This
broadening is explained as due to the light intensity decay
in course of propagation along the Bragg direction
{\em before the first} and {\em after the last} scattering events.
We modified the CBS theory to incorporate the attenuation that
results from the photonic band structure of the medium.
Using the modified theory we extract from our CBS data the 
light mean free path and Bragg attenuation length at different
$\left[hkl\right]$. Our study shows that CBS measurements are a unique
experimental technique to explore photonic crystals with disorder,
when other spectroscopical methods become ambiguous 
due to disorder-induced broadening.

\end{abstract}

\newpage

\section{Introduction}

Expreimental studies of 
coherent backscattering (CBS) of light from  disordered
media have been completed
by many groups  during the last decade.$^{1-8}$ 
CBS measurements  are of great interest
because they have allowed to test the physics of weak
localization, which was first uncovered in  electronic
transport,\cite{langer}
when applied to light waves  and also to determine the light
mean-free-path, $l^{\ast}$ in  random media.
In CBS measurements the normalized reflected light intensity
(also called the albedo, $\alpha$) is monitored versus the deviation
angle, $\delta \theta$ from the backscattering direction
of the incident laser beam. In the multiple scattering regime
interference effects are  averaged out except in a small
angular range around $\delta\theta=0$, where constructive interference
originating from reciprocal light paths enhance the diffusive
reflection from the sample. As a result a coherent scattering
intensity cone is formed in the albedo, $\alpha(\delta\theta)$
for small $\delta\theta$. From the full width, $\Delta\theta$
of the CBS cone at quarter maximum, $l^{\ast}$ is obtained
using the relation $\Delta\theta=\lambda_L/2\pi l^{\ast}$,
where $\lambda_L$ is the impinging laser wavelengh.

The first CBS
measurements were performed on suspensions of polystyrene
spheres.\cite{kuga84,albada85} Since the suspension particles were very
small in diameter, their Brownian motion  provided a natural
ensemble averaging. For solid materials, however CBS
usually shows a noisy but repeatable mesoscopic 
pattern.\cite{berkov94} Under these conditions the CBS
cone around $\delta\theta=0$ can  be obtained  by applying an elaborate
ensemble-averaging
technique in order to suppress the dominant mesoscopic
fluctuations. The averaging techniques  employed to get the
cone include 
spot variation,\cite{etemad86} tilt angle variation,\cite{etemad86} 
and sample spin-rotation.\cite{kaveh86} 
All the above experimental 
studies$^{1-8}$
were performed on completely disordered, isotropic and
homogeneous media.

Photonic crystals have recently attracted a lot of
attention\cite{Photo,books,Zakhidov,Frolov} because of potential
optical device applications. These applications pertain to photonic
crystals with
{\em complete} band gaps {\em i.e.} with no propagating solutions of
the Maxwell equations within a certain frequency range. However, most
of the available three dimensional 
photonic crystals nowadays have {\em incomplete} 
gaps in
their photonic band structure, with light propagation forbidden only
along certain directions.   
Along these directions and for frequencies within the gap
the light intensity decays
with a certain decrement $L_B$-that is the Bragg attenuation
length.  In the ideal case 
(when  disorder is negligibly small) the length
$L_B$ can be extracted from spectroscopic 
reflectivity measurements.
Since $L_B$ is directly related to the gap width, then its value can be
 deduced 
from the angular (or spectral) width of the Bragg peak in 
reflectivity. However, in the more realistic situation 
  disorder is
present in the sample and this broadens the Bragg 
band. We also note that even if $L_B\ll l^{\ast}$ 
 broadening can be substantial
due to the sensitivity of the Bragg diffraction to the
properties of the surface boundary.  
Under these conditions a separate determination
of $L_B$ and $l^{\ast}$ is a challenge.

In general, despite the conceptual elegance of the CBS technique, the
main quantitative information it has provided$^{2-8}$ so far is 
$l^{\ast}$ value in {\em macroscopically homogeneous}
and isotropic disordered media.
We note
that $l^{\ast}$ value in such media
 can be also  determined from more conventional
measurements of the transmission versus thickness\cite{genack87}
that do not rely on the interference of clockwise and anticlockwise
scattering paths.

In the present work we show that for optically  {\em periodic}
structures {\em with disorder} the CBS measurements provide
a unique opportunity to separate the two effects, namely
Bragg attenuation and light diffusion lengths. This reveals
the tremendous potential of the CBS technique as compared 
to the more conventional optical spectroscopies.

\section{Experimental Results}

For our studies of periodic structures
we have chosen  the synthetic $SiO_2$ opals 
that are actually three-dimensional photonic crystals.\cite{Zakhidov}
The synthetic opal single crystals were fabricated  from a 
colloidal suspension of
mono-dispersed silica ($SiO_2$) spheres with a mean diameter,
 $\cal D$=$295 nm$ and
dispersion in $\cal D$ of about $4\%$. In contrast to the natural 
gem opals that exist in the market the space
between $SiO_2$ spheres in artificial opals is filled  with air. 
A sintering process leads to
the formation of inter-sphere necks that mechanically 
connect neighboring
silica spheres in a face-centered-cubic (fcc) lattice. 
The opals with air-filled voids are chalk-like diffuse 
light scatterers that show some
iridescence due to Bragg diffraction from their crystal planes. 
The iridescence
is dramatically enhanced, however if the voids are filled with 
 an index matching liquid.

In opals there exist photonic gaps, $\Delta \nu_{hkl}$,  in the
frequency spectrum around each $\nu_{hkl}$, where $(hkl)$ are the Miller indices of the
crystal planes. The  central frequencies, $\nu_{hkl}$,
 are determined
 by the Bragg condition: $\lambda_B=c\nu_{hkl}^{-1} =
2n_{eff}d_{hkl}$, 
where $d_{hkl}$ is the
inter-planar spacing, $n_{eff}$ 
is the opal effective index of refraction ($n_{eff} =1.376$
 for air-filled opals) and $c$ is the speed of light. 
In the absence of disorder the stop band width, $\Delta\nu_B$ is
related  to the Bragg attenuation length, $L_B$, by the relation
$\Delta\nu_B/\nu_B = 2d/\pi L_B$ (with implied $hkl$ indices),
where $\nu_B=c/\lambda_B$. As mentioned above, however disorder
in opals dramatically broadens $\Delta\nu_B$ so that this relation
is no longer valid.  

For our CBS measurements, we cut a centimeter-size opal single crystal
into planes
orthogonal to several $\left[hkl\right]$ directions
including the fast growing $\left[111\right]$ direction
and a prominent $\left[220\right]$ direction. We note
that the $\left(200\right)$ planes in opals are rather disordered.

We used a standard experimental set-up\cite{etemad86,kaveh86}
 with angular resolution in $\delta\theta$ of less than
$1~ mrad$. The coherent light beam was directed from continuous wave
lasers such as $Ar^+$ and $HeNe$, as well as various well-collimated
semiconductor lasers. 
The laser beam was linearly polarized with transverse
electric field ($TE$) polarization. 
The reflected intensity from the opal
$\left(hkl\right)$ surface with $TE$ polarization was 
measured versus $\delta\theta$ using a silicon
detector and phase-sensitive techniques. The incident beam ($5 mm$ in
diameter) was directed at various angles, $\theta$   between 
$-30^{\circ}$ to $70^{\circ}$
relative to the surface
normal, and $\delta\theta$ was varied within $200~ mrad$.
To obtain specle-free CBS cones we performed configuration averaging
using three different methods. These include different illuminated
areas of the opal surface, different inclination angles 
$\varphi < 2^{\circ}$ of the illuminated surface with respect to
the laser beam, and sample rotation about the normal direction to the
crystal surface.

In Fig. 1 we show typical CBS cones measured 
at various incident angles $\theta$
with respect to $\left[111\right]$
that we obtained using an $Ar^+$ laser
at $\lambda_L = 515 nm$.
The CBS cones are apparent in all cases
and their width, $\Delta\theta$ at quarter maximum versus $\theta$ 
is summarized in Fig. 2. Except
for a small $\theta$ interval, $\gamma$ (about $5^{\circ}$)
around $\theta =42^{\circ}$, 
$\Delta\theta$ monotonously increases with $\theta$. 
Within the interval $\gamma$ around $\theta =42^{\circ}$, however  
there is an anomalous increase in $\Delta \theta$ of about a
factor of two (Fig. 2).

 Insight into the origin of the observed 
 anomaly in $\Delta\theta$ ($\theta$)
may be obtained from Fig.~3 in which the angle-dependent 
reflectivity spectra, $R(\lambda)$, from the $\left(111\right)$
opal surface are shown.
It is seen that for $\lambda_L=515 nm$ the angle 
$\theta=\theta_B=42^{\circ}$,
at which the anomaly in the CBS cones occurs, 
in fact corresponds to the 
Bragg conditions.

Another evidence that the 
CBS anomaly is related to the  Bragg conditions 
can be  observed from the obtained CBS cones at $\theta$
around $\theta_B = 0$ (Fig. 4).
These CBS cones
 were measured using a $HeNe$ laser at $\lambda_L = 633nm$
that is close to $\lambda_B$ at $\theta = 0$ (Fig. 3, inset). 
Fig. 4 inset shows that
$\Delta\theta$ for CBS cones measured at normal incidence 
stays approximately constant
at about $17 mrad$ for laser wavelengths outside the Bragg conditions. 
However,
for CBS measured at $\theta\approx\theta_B=0$, 
$\Delta\theta$ in the CBS cone increases by a
factor of $\approx 4$. Also the incident angle interval 
($\gamma\approx 20^{\circ}$, Fig. 4) in which the
CBS anomalies are observed with laser illumination at 
$\lambda_L = 633 nm$ is much
larger than that obtained at $\lambda_L = 515 nm$ ($\gamma\approx 
5^{\circ}$, Fig. 2). Both observations, namely the anomaly in 
$\Delta\theta$ ($\theta$) at $\theta=\theta_B$ and strengthening
of the anomaly at $\theta_B\approx 0$ require an explanation.
A possible model is given in the next Section.

\section{Discussion}
\label{disc}

The  Bragg-related anomalies of the CBS cones show that light
transport and localization is substantially affected by the
$\left[111\right]$ photonic band gap in the opal, where light becomes
more localized. To understand these results  we first recall the most
transparent derivation of the CBS cones in the literature
and apply it 
 to laser frequencies outside the photonic gap.\cite{akk86,wolf88}
We then modify the theory to interpret the anomalous CBS
results at intra-gap laser frequencies at
$\theta\approx\theta_B$.

The coherent scattering albedo, $\alpha_c(\theta,\theta^{\prime})$, at
an angle $\theta^{\prime}$ from a semi-infinite slab at $z~=~0$ due to
an incident laser beam at an angle $\theta$ is determined by the first
and last light scattering events occurring at the spatial points $\bf
r$ and $\bf r^{\prime}$, respectively, and by the light diffusion
process that occurs between these events. If $z$ and $z^{\prime}$ are
the projections of $\bf{r}$ and $\bf{r}^{\prime}$ on the $z$-axis, and
$\rho$ is the projection of $(\bf{r}-\bf{r}^{\prime})$ on the sample
interface plane, then the coherent part of  albedo 
is given by the expression
\begin{equation}
\label{albedo}
\alpha_c(\theta,\theta^{\prime})\propto\int_0^{\infty}dzdz^{\prime}
\int d^2\rho {\cal P}_{\phi}(z)\left\{\cos\left[k_{\perp}\rho+
k(\cos\phi -\cos\phi^{\prime})(z-z^{\prime})\right]\right\}
\Pi(\rho,z,z^{\prime}){\cal P}_{\phi^{\prime}}(z^{\prime}).
\end{equation}
Here $k = 2\pi n_{eff}/\lambda_L$ and $k_{\perp} = k(\sin\phi -
\sin\phi^{\prime})$, where the angles $\phi$ and $\phi^{\prime}$ are,
respectively, the propagation directions of the incoming and outgoing
light {\em inside} the slab.  These angles are related to $\theta$ and
$\theta^{\prime}$ through the Snell law $n_{eff}\sin\phi =\sin\theta$.
In Eq. (\ref{albedo}) ${\cal P}_{\phi}(z)$ is the probability density
for the incident light to propagate up to the depth $z$ {\em before} the
occurence of the
{\em first} scattering event.  Analogously, ${\cal
P}_{\phi^{\prime}}(z^{\prime})$ is the surviving probability density
{\em after} the occurence of the
{\em last} scattering event.  ${\cal
P}_{\phi^{\prime}}(z^{\prime})$
has the same form as
${\cal P}_{\phi}(z)$, where $\phi$, $z$ are replaced by
$\phi^{\prime}$, $z^{\prime}$ respectively. 
Since ${\cal P}_{\phi}(z)$ satisfies
the Poisson statistics, it is given by ${\cal P}_{\phi}(z)=
(l^{\ast}\cos\phi)^{-1}\exp\left(- z/l^{\ast}cos\phi\right)$.  The
propagator $\Pi(\rho,z,z^{\prime})$ in Eq. (\ref{albedo}) describes
the diffusive motion of light between the first and last scattering
events.  To take into account  the fact that diffusion occurs in the
presence of the boundary $z=0$, it is customary\cite{akk86,wolf88} to
choose the propagator $\Pi$ in the form
\begin{equation}
\label{Pi}
\Pi(\rho,z,z^{\prime})=\Pi_0(\rho,z,z^{\prime})-
\Pi_0(\rho,z,z^{\prime\ast}),
\end{equation} 
where $\Pi_0(\rho,z,z^{\prime}) = \left[4\pi
D\left(\left(z-z^{\prime}\right)^2+ \rho^2\right)^{1/2}\right]^{-1}$
is the diffusion propagator in an unbounded medium; $D=l^{\ast}c/3$ is
the light diffusion coefficient, and $z^{\prime\ast}$ is the mirror
image of $z^{\prime}$ with respect to the plane $z=-z_0$, where
$z_0=0.7l^{\ast}$.\cite{akk86,wolf88} The choice (\ref{Pi}) of the
propagator $\Pi$ insures that the boundary conditions
$\Pi(\rho,-z_0,z^{\prime})=\Pi(\rho,z,-z_0)=0$ are obeyed.  The
evaluation of integrals in Eq. (\ref{albedo}) for small
$\delta\theta=\theta - \theta^{\prime}$ yields\cite{akk86,wolf88}
\begin{equation}
\label{alpha}
\alpha_c\left(\delta\theta\right)\propto\left(\frac{\cos\phi}
{1+k_{\perp}l^{\ast}\cos\phi}\right)^2\left(\frac{1-e^{-2k_{\perp}z_0}}
{k_{\perp}l^{\ast}}+\cos\phi\right),
\end{equation} 
where $\delta\theta$ enters into the r.h.s. of Eq. (\ref{alpha})
through $k_{\perp}=2\pi\cos\theta \delta\theta/\lambda_L$ and $\theta$
and $\phi$ are related through  Snell's law.
 Eq. (\ref{alpha}) contains only a single free
parameter, namely $l^{\ast}$.

We used Eq. (\ref{alpha}) to fit the CBS cones in
$\alpha_c(\delta\theta)$ at various $\theta$ outside the
anomaly angle interval
$\gamma$ around $\theta = \theta_B$. The excellent fits shown in
Fig. 1
 for $\theta = 10^{\circ},34^{\circ}$ and $60^{\circ}$ were
obtained with a single $l^{\ast} = 7.2 \mu m$.  From this value for
the optical mean-free-path length and Eq. (\ref{alpha}), we calculated
the width of the albedo for the CBS cone at quarter maximum,
$\Delta\theta$, versus $\theta$, as shown in Fig. 2.
Except near $\theta_B$ the overall
agreement between the calculated and the experimentally determined
$\Delta\theta$ is excellent  showing that
light transport is quite
ordinary for frequencies outside the photonic gap.

For describing the CBS process at $\theta\approx\theta_B$ and
correspondingly at $\phi\approx\phi_B$ we note that the diffusive
propagation of light between the first and last scattering events that
is
given by $\Pi(\rho,z,z^{\prime})$ in Eq. (\ref{albedo})  is affected
very little by the Bragg diffraction. The reason for this is the small
angle interval, $\beta(\theta_B)$, around $\theta_B$ where  Bragg
condition is satisfied. An upper bound
 estimate for $\beta(\theta_B)$ can be
obtained by relating it to the spectral width of the Bragg gap
$\delta\lambda_B$, where
$\delta\lambda_B/\lambda_L =\beta(\theta_B)\sin2\theta_B/
2\left(n_{eff}^2-\sin^2\theta_B\right)$. If we choose for
$\delta\lambda_B/\lambda_L$ the measured value of the reflectivity peak
$\delta\lambda_B/\lambda_L=0.09$, then for $\lambda_L=515nm$ and
$\theta_B\approx 42^{\circ}$ we calculate $\beta$ to be $0.25$~radian.
This forms a stereo-angle of about $0.07$ stereo-radian around the
Bragg direction that is only about $0.5\%$ of the complete $4\pi$
stereo-angle.  Therefore, even if the light beam is incident at
$\theta = \theta_B$, it is diverted after the first scattering event
and is then followed by an almost regular diffusion;
this is true since the
probability for light to propagate close to $\theta_B$  again 
is very
small. This permits the generalization of the CBS theory to photonic
crystals with relatively small stop bands, simply by only modifying 
the probability densities ${\cal P}_{\phi}(z)$ and ${\cal
P}_{\phi^{\prime}}(z^{\prime})$ in Eq. (\ref{albedo}).  For
$\theta\approx\theta_B$, light intensity decays as
$\exp\left(-z/L_B\right)$ due to the Bragg diffraction from the
photonic crystal planes. Therefore ${\cal P}_{\phi}(z)$ can be written
at $\theta = \theta_B$ as a product of the surviving probabilities:
$\exp\left(- z/l^{\ast}cos\phi\right)$ and the Bragg attenuation,
$\exp\left(-z/L_B\right)$.  Equivalently, we may represent ${\cal
P}_{\phi}(z)$ at $\theta = \theta_B$ as a single exponent,
$\exp\left(-z/l^{\ast}_{eff} \cos\phi_B\right)$, where
$\left(l_{eff}^{\ast}\left(\phi_B\right)\right)^{-1}=\left(l^{\ast}\right)^{-1}+
\cos\phi_B\left(L_B\right)^{-1}$.  This expression shows that
Eq. (\ref{alpha}) describing the CBS cone is still valid at
$\theta=\theta_B$, but with a smaller effective mean-free-path,
$l^{\ast}_{eff}$.  A smaller $l^{\ast}_{eff}$ leads to a broader CBS
cone at $\theta=\theta_B$, as indeed seen in Fig. 2 for $\theta_B\approx
42^{\circ}$.

If $\theta$ and $\theta^{\prime}$ are close, but not equal to
$\theta_B$, then the Bragg attenuation becomes weaker. 
It can be shown that
this effect is taken into account by the following modification of
$l^{\ast}_{eff}(\phi)$
\begin{equation}
\label{eff}
\frac{1}{l^{\ast}_{eff}\left(\phi\right)}=\frac{1}{l^{\ast}}
+\frac{\cos\phi_B}{L_B}\left[1-\frac{\pi^2L_B^2\sin^2\left(2\theta_B\right)}
{\lambda_L^2\left(n_{eff}^2-\sin^2\theta_B\right)}
\left(\theta-\theta_B\right)^2\right]^{1/2}.
\end{equation}
Since the argument under the square root in Eq. (\ref{eff}) must be
positive, we use this constraint to estimate the angle interval
$\gamma$ around $\theta_B$ where the CBS anomalies due to the photonic
crystal are observed.  This leads to the expression: $\gamma(\theta_B)
= \lambda_L^2/\pi L_B d\sin\left(2\theta_B\right)$.  Fig. 2
shows that
an excellent fit is obtained for both $\Delta\theta$ enhancement and
$\gamma$, using Eqs. (\ref{alpha}) and (\ref{eff}) with $L_B =5.1 \mu
m$ and the previously obtained $l^{\ast} = 7.2 \mu m$.  Using these
values of $l^{\ast}$ and $L_B$ along $\left[111\right]$ we evaluated
$\gamma(42^{\circ})=4^{\circ}$, which agrees well with the data of
Fig. 2.  We also calculated the ``pure'' Bragg band
$\delta\lambda_B$ to be
$\delta\lambda_B/\lambda_L \approx 0.03$. We  note, that this value is 3
times smaller than $\delta\lambda/\lambda_L \approx 0.09$ extracted
from the Bragg reflectivity band in Fig. 3.   This 
illustrates  the ambiguity in determining  the photonic
band structure form the more conventional  reflectivity 
spectroscopy.

For $\theta \approx 0$ {\em i.e.} 
close to normal incidence and $\lambda_L \approx \lambda_B$,
Eq. (\ref{eff}) is no longer valid, since $\vert \theta -\theta_B\vert
\sim \theta_B$.  In this case it can be shown that  the expression
for the effective light  mean-free-path is given by
\begin{equation}
\label{eff1}
\frac{1}{l^{\ast}_{eff}\left(\phi\right)}=\frac{1}{l^{\ast}}
+\frac{1}{L_B}\left[1-\left(\frac{\pi L_B \theta^2}
{\lambda_L n_{eff}}\right)^2
\right]^{1/2}.
\end{equation}
Again the constraint that the argument under the square root 
is positive now
yields $\gamma(0^{\circ})=26^{\circ}$. This value is much larger than
the obtained $\gamma(42^{\circ})=4^{\circ}$ for $\lambda_L = 515 nm$.
However this much larger value is consistent with our measurements
at $\lambda_L=633nm$
(Fig. 4).  Moreover, from Eq. (\ref{eff1}) we calculated 
theoretical curve $\Delta\theta$ $(\theta)$ shown in Fig. 4.
 This
curve fits very well the width enhancement of the CBS cone at
$\theta_B = 0$, when $\theta$ is changed for a fixed $\lambda_L$
(Fig. 4 for $\lambda_L = \lambda_B = 633 nm$).  Eq. (\ref{eff1}) is
also consistent with the obtained results when $\lambda_L$ is changed
for a fixed $\theta =0$ (Fig. 4 inset).  The agreements
obtained validate both our
modified CBS model and the extracted optical parameters for the opal
photonic crystal.  We again
 emphasize that consistent values of $L_B$ were
inferred from both the magnitude $\Delta \theta$ $(\theta_B)$ and the
width $\gamma (\theta_B)$ of the CBS anomaly.

The widths of CBS cones from $\left(220\right)$ planes
measured with $\lambda_L=365nm$ are shown in
Fig.~5, where $\Delta\theta$ is plotted versus the incident
angle $\theta$ measured from $\left[220\right]$ direction.
For  $\lambda_L=365nm$ the Bragg CBS anomaly 
in $\Delta\theta$ occurs at 
$\theta=\theta_B=
\cos^{-1}\left(\lambda_L/2d_{220}n_{eff}\right)
\approx 20^{\circ}$, and is clearly seen in Fig.~~5.
Fitting this CBS anomaly in a similar way as that 
 at $\theta=42^{\circ}$ in  Fig. 2, we deduced a 
Bragg length, $L_B=4.3\mu m$ 
for the photonic gap in $\left[220\right]$ direction.
In our fitting we used the value of mean-free-path,
$l^{\ast}=4.6\mu m$ that was extracted from the widths
of the CBS cones for incident light beams close to the normal
direction to 
the $\left(220\right)$ planes.
We note that, in addition to $L_B$ along $\left[220\right]$,
$l^{\ast}$ is also different from that extracted before 
from measurements along 
$\left[111\right]$.

\section{Microscopic origin of the scatterers}




Apart from the Bragg-related anomalies we note that
 three {\em qualitative} features of the data presented
 above  indicate that
the description of  CBS from opal photonic crystal within the
conventional model of a medium with randomly positioned
point-like scatterers is not completely adequate. These features
are:

({\em i}) The Rayleigh scattering
 law $l^{\ast}\propto \lambda_L^{-4}$ that is
characterisitc for point-like scatterers does not hold in opals.
This fact becomes evident from  Fig. 4 inset;  
except for a narrow $\lambda_L$ interval near 
$\lambda_L\approx \lambda_B$,
the
CBS cone width $\Delta\theta$ remains approximately  unchanged 
with $\lambda_L$ for 
$\lambda_L$ interval between $800nm$ to $360nm$.

({\em ii}) The value of $l^{\ast}=7.2\mu m$ for 
$\left[111\right]$ direction   appreciably  exceeds
$l^{\ast}=4.6 \mu m$ for $\left[220\right]$ direction.

({\em iii}) The cone width $\Delta\theta$ increases with $\theta$
for CBS along $\left[111\right]$ (Fig. 2) but shows no dependence 
on $\theta$ for CBS along $\left[220\right]$ (Fig. 5).

The question that we would like to address here is whether
these CBS features can help in pinpointing  the
size, shape, and orientation
of the generic defect scatterers in opal.
 
Firstly, independence of $l^{\ast}$ on $\lambda_L$
suggests that the size of the defects is larger than $\lambda_L$. 
On the other hand, defects with size ${\cal L} \gg \lambda_L$ 
scatter most of the light in the
forward direction. More precisely, the scattered light intensity is
restricted to a cone with  aperture $\sim \lambda_L/\cal L\ll \pi$.
As a direct consequence of this almost forward scattering, 
we should
have $l_{tr}^{\ast}\gg l^{\ast}$, where  $l_{tr}^{\ast}$ is the
{\em transport} mean-free-path  that determines the diffusion
coefficient ($D=cl_{tr}^{\ast}/3$) in this case. 

The conclusion that $l_{tr}^{\ast}$ is much bigger than $l^{\ast}$
in opal has some
dramatic consequencies.
 As  can be seen from the general expression 
for the coherent albedo [Eq. (\ref{albedo})], the dependence 
$\alpha_c(\delta\theta)$
is determined by two factors: the surviving probabilities 
${\cal P}_{\phi}$, 
${\cal P}_{\phi^{\prime}}$ and the propagator $\Pi$. 
For point-like scatterers both ingredients change within
 the same characteristic
spatial scale, namely $l^{\ast}$. However if 
 $l_{tr}^{\ast}\gg l^{\ast}$,
then the scale of the propagator that describes the
diffusion process is  $l_{tr}^{\ast}$, whereas the scale of 
${\cal P}_{\phi}$, 
${\cal P}_{\phi^{\prime}}$ is still $l^{\ast}$. This 
would make the cone width $\Delta\theta\sim \lambda_L/l_{tr}^{\ast}$
rather than $\Delta\theta\sim \lambda_L/l^{\ast}$ as in the usual case.
Consequently $\Delta\theta$ would be
 {\em independent} of the incident angle $\theta$,
and, correspondigly insensitive to the Bragg anomaly,
contrary to the data.

A possible resolution of this apparent contradiction with the
experiment that we can suggest here is to assume
that the scatterers are
{\em highly anisotropic}, such as disk-like with radius 
$\sim {\cal L} \gg \lambda_L$ and thickness $\sim \lambda_L$.
Moreover, in order to insure that CBS is different from the
$\left(111\right)$ and $\left(220\right)$ opal surfaces
as found experimentally, we
have to assume that the disks are not  oriented  randomly,
but rather their orientation is  correlated. Although we did not extend 
the quantitative theory of CBS to this situation, intuitively
it seems  likely that the disks normals be parallel
to the $\left[111\right]$ direction.  For such an 
orientation the light would diffuse easier {\em along} the 
$\left(111\right)$ planes. Under these conditions of strongly
anisotropic diffusion, longer paths do not necessarily lead
to narrower cones as in the case of isotropic scatterers.
Since CBS is essentially the ``structure factor'' 
of the propagation {\em along the boundary},\cite{akk86}
then the easier
diffusion along $\left(111\right)$
 would lead to narrower cones, {\em i.e.}
to $\Delta\theta_{\left[111\right]} < \Delta\theta_{\left[220\right]}$,
in agreement with the experimental results (Figs.~2 and 5).

A possible realization of the geometry and arrangement
of scatterers as
conjectured above can be provided by stalking faults defects. 
 Such  defects can be easily created
along the $\left[111\right]$ opal fastest growing direction. Also, 
this type 
of defects has been recently shown to dominate disorder
in colloidal crystals,\cite{bertone99} and we note that synthetic
opals grow from silica colloids.

\section{Conclusions}

In conclusion, we have demonstrated that CBS is a powerful technique
for investigating photonic crystals with incomplete photonic band
gaps. These measurements show that the remarkable universality of CBS
characteristics for diverse disordered materials is not valid for
photonic crystals at near Bragg conditions.
Using the major effects on light transport near Bragg conditions and
their influence on CBS cones, we deduce the values
of several fundamental parameters of an opal
photonic crystal.  Remarkably,  despite the existence of
substantial scattering due to disorder, nevertheless
 we showed in this work 
that the Bragg
attenuation length and, thus the photonic band gap width of an ``ideal''
crystal along $\left[111\right]$, as well as light mean-free-path can
be readily extracted from the CBS measurements, even if they are
comparable in magnitude (unlike the situation in 
Ref. \onlinecite{bertone99}).

The reason
why for $L_B\lesssim l^{\ast}$ disorder is unable to completely mask
the contribution of Bragg scattering to the CBS
 inside the photonic crystal, is that  at Bragg
conditions the CBS
involves $\em free$ light propagation {\em before the
first} and {\em after the last} scattering events, when the existing
disorder is {\em not yet relevant}.  This makes the CBS measurements a
much more sensitive tool for characterizing the inherent band gap
features for disordered photonic crystals as compared to the 
more conventional measurements of optical reflectivity.

The only other study of  CBS from photonic crystals that we know of 
was recently reported in Ref. \onlinecite{Vos}. 
The CBS measurements in that work  were performed on photonic 
crystals based on
 polystyrene spheres and also on crystals composed of 
air spheres in a $TiO_2$ matrix. In contrast to the present study
that has been performed  on a {\em single}
opal sample where the stop-band was ``swept''
by changing the  angle of incindence, $\theta$; in  
 Ref. \onlinecite{Vos}, on the contrary the incident angle was kept
close to  normal. Under this condition, in  order to
explore the CBS within the stop-band frequency range in the
polystyrene
opals, the authors  measured CBS on a {\em set} of samples
with different sphere diameters using 
three different wavelenghts for the  incident
light beam.
Also the band edge frequencies   were 
determined from  {\em reflectivity} measurements,
rather than by CBS as in our work. Even under these assumptions,
the authors failed to measure  CBS cones 
{\em inside} the stop-band.\cite{Vos}

For air-sphere $TiO_2$ crystal, however,
 the authors\cite{Vos} used a setup 
allowing continuous tuning of $\lambda_L$. They observed that
between the red and  blue edges of the stop-band,
$\Delta\theta$ {\em decreases} by  a factor of $\sim 2$. 
 This observation, which actually contradicts our own study,
 was explained on the basis of a previous theoretical
work.\cite{Lagen89}
It was claimed that for  light frequencies 
above  the
lower band-edge  the electromagnetic wave, in course
of the diffusion process,  
can occasionally ``hit'' the boundary interface at
the Bragg angle $\theta=\theta_B$; thus rather than
exiting the sample, it experiences instead the
Bragg diffraction to the interior. As a result of such 
events the diffusive
trajectories become longer (on average) leading to the 
narrowing of the CBS cone.

In light of the CBS model that was outlined in Section III,
the above explanation corresponds to the modification of the propagator
$\Pi(\rho,z,z^{\prime})$ in Eq. (\ref{albedo}).
 In contrast, our analysis
suggests that $\Pi$ is
not appreciably affected by the Bragg conditions, whereas the prime 
effect of the photonic band structure is the modification
of the surviving probability densities ${\cal P}_{\phi}(z)$ and
${\cal P}_{\phi^{\prime}}(z^{\prime})$ in Eq. (\ref{albedo}).
 The  apparent
contradiction
between our results and those of Ref. \onlinecite{Vos} can be probably
 accounted for by the difference in the 
``strenghts''of the Bragg diffraction process, or $L_B$
in synthetic $SiO_2$ opals
and  air-sphere crystals in $TiO_2$. We note, however that  
the relative width of the Bragg reflectivity peaks in both samples
is comparable $\delta\lambda_B/\lambda \sim 0.1$.

\acknowledgements

This work was supported by NSF grant DMR 973280, the 
Petroleum Research Fund under grant ACS-PRF \#34302-AC6  and
the Army Research Office.


\begin{figure}\label{fig1}
\caption{Coherent backscattering cones in the albedo,$\alpha$
at various incident
angles $\theta$  from $\left[111\right]$ measured at 
$\lambda_L = 515 nm$; $\theta=42^{\circ}$ 
is the Bragg angle at $\lambda_L=515 nm$. Fits to the experimental data
are also shown, using Eq. (\ref{alpha}) with $l^{\ast}$
 as a free parameter.} 
\end{figure}

\begin{figure} \label{fig2}
\caption{The
width  $\Delta\theta$, at quarter maximum of
 the CBS cones obtained at $\lambda_L=515 nm$     
 versus the
incident angle, $\theta$. The full line is calculated using 
Eq. (\ref{alpha}) with $l^{\ast} = 7.2 \mu m$; the dashed-broken line 
around $\theta_B = 42^{\circ}$ that describes the
Bragg-related CBS anomaly, is calculated using Eqs. (\ref{alpha}) and 
(\ref{eff}) with $l^{\ast} =7.2 \mu m$ and $L_B = 5.1 \mu m$.}
\end{figure}

\begin{figure}\label{fig3}
\caption{Normalized reflectivity spectra for various Bragg
angles $\theta$ with respect to the opal $\left[111\right]$
fast growing direction. 
The inset shows the wavelength of the Bragg band, $\lambda_B$
vs. $\theta$; the full line is a fit using the relation:  
$\lambda_B(\theta)=2d_{111}\left(n_{eff}^2-\sin^2\theta\right)^{1/2}$.}
\end{figure}

\begin{figure}\label{fig4}
\caption{The width at quarter maximum, $\Delta\theta$ 
of CBS cones measured at various $\theta$ around $\theta =0$
using $\lambda_L = 633 nm$, for which $\theta_B\approx 0$. 
The dashed line is
calculated using  Eqs. (\ref{alpha}) and (\ref{eff1}). The inset shows
$\Delta\theta$ of CBS cones measured at $\theta=0$  
using various laser 
frequencies. The dashed line is a theoretical 
fit to the data.} 
\end{figure}

\begin{figure}\label{fig5}
\caption{The
width  $\Delta\theta$, at quarter maximum of
 the CBS cones obtained at $\lambda_L=365 nm$     
 versus the
incident angle, $\theta$. The dashed-broken line 
around $\theta_B = 20^{\circ}$ that describes the
Bragg-related CBS anomaly, is calculated using Eqs. (\ref{alpha}) and 
(\ref{eff}) with $l^{\ast} =4.6 \mu m$ and $L_B = 4.3 \mu m$.}
\end{figure}

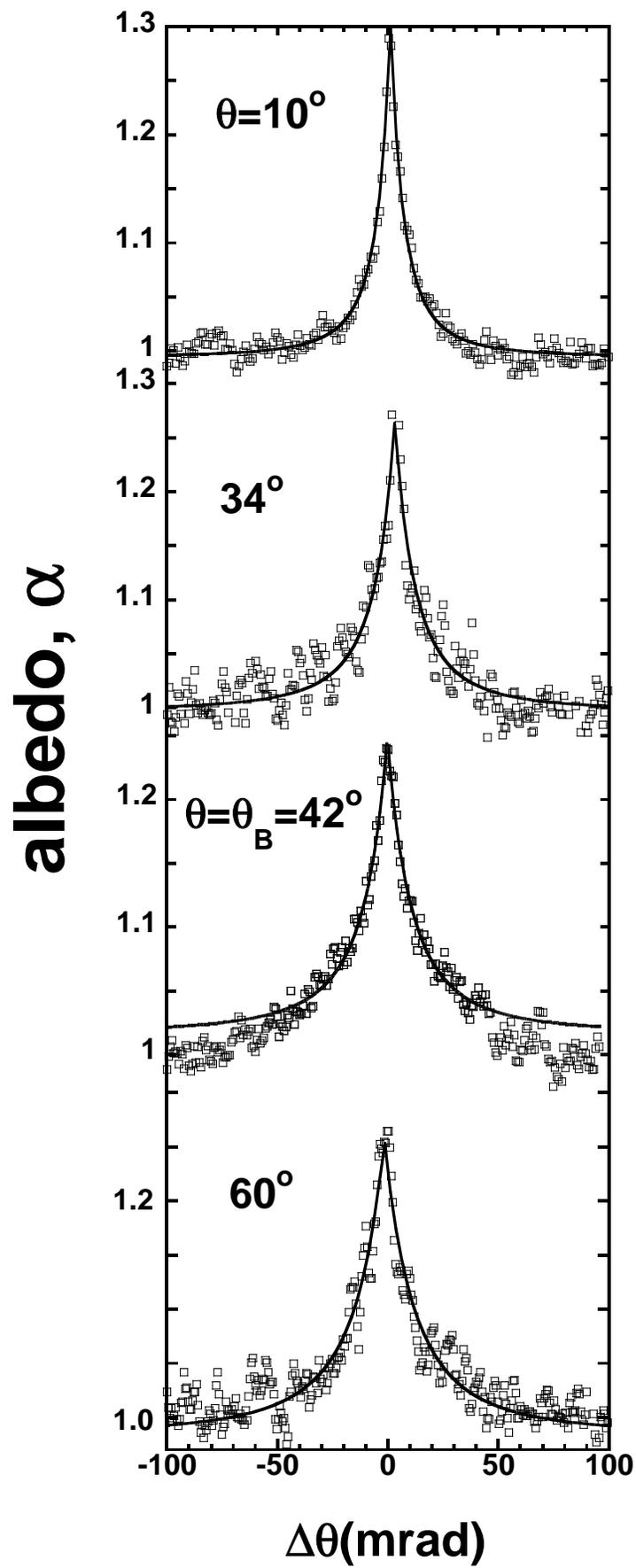
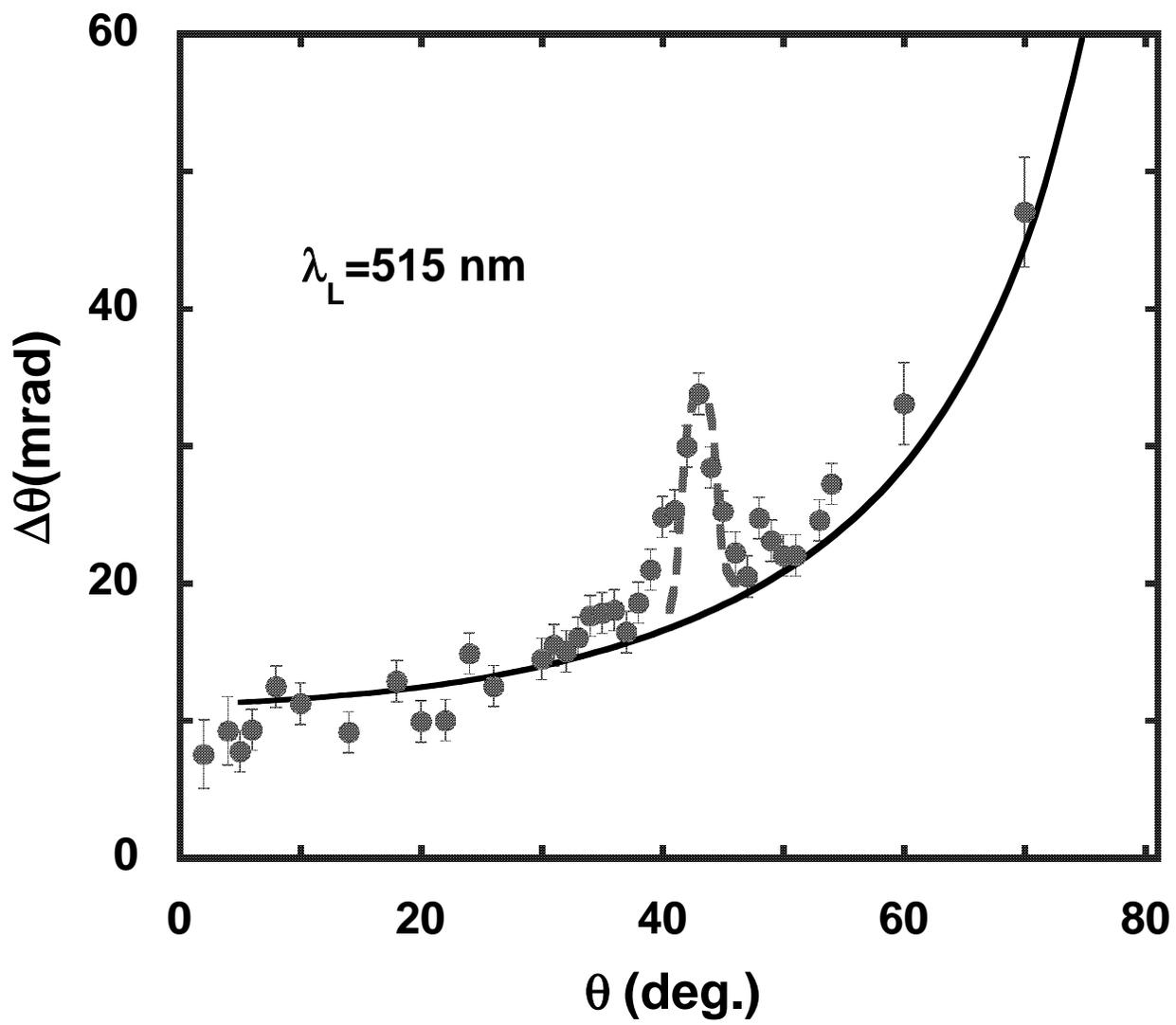
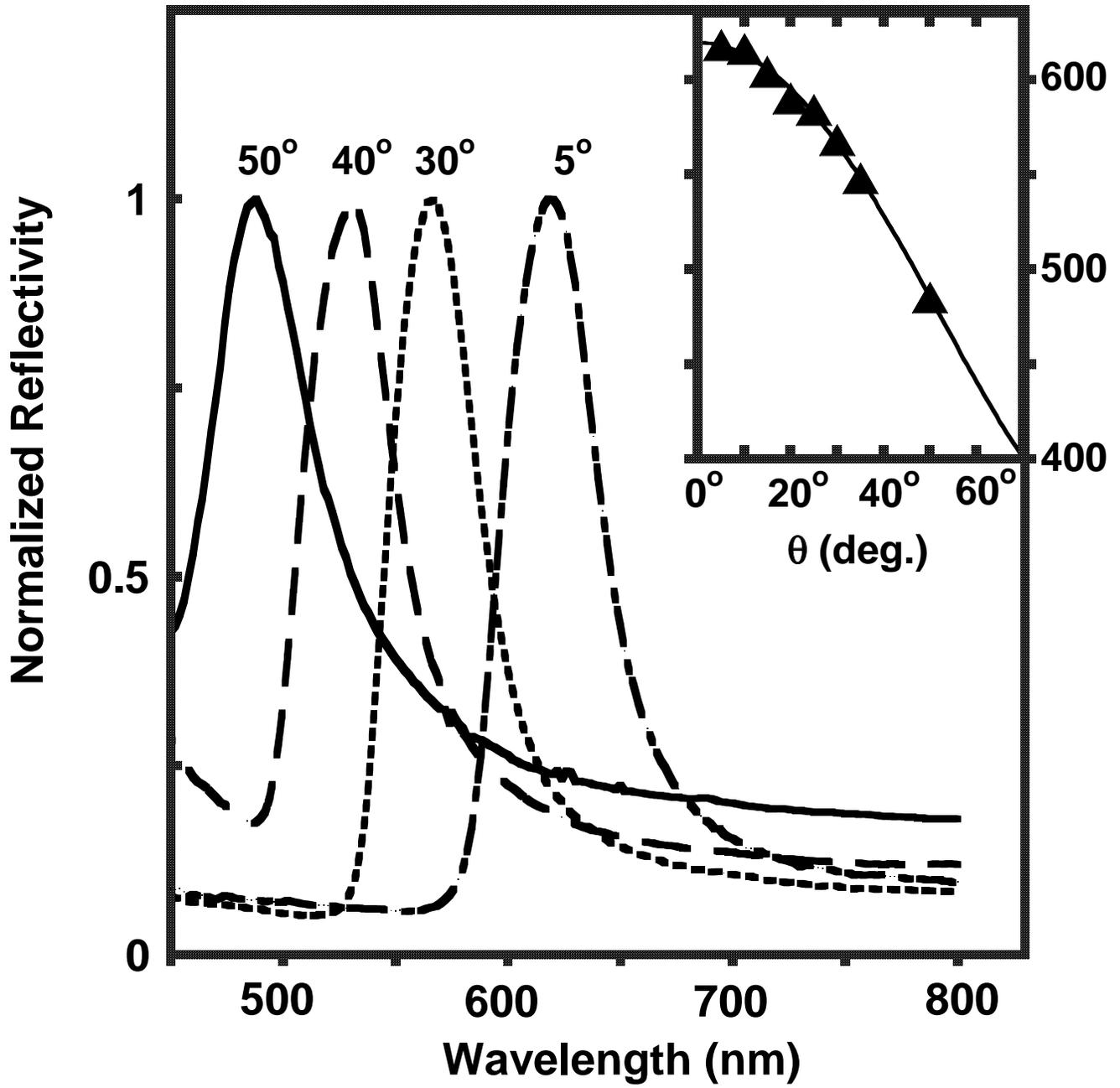
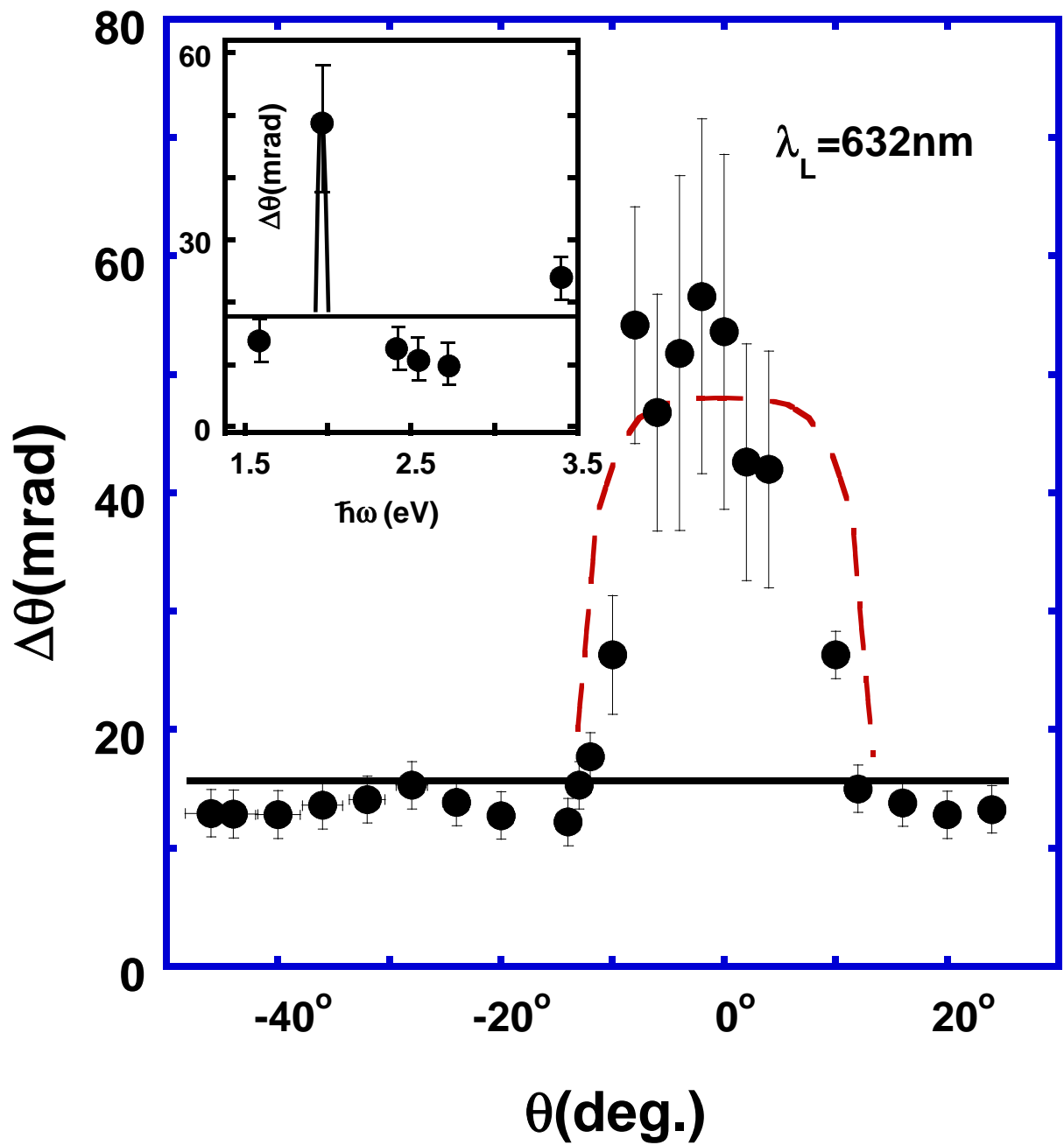
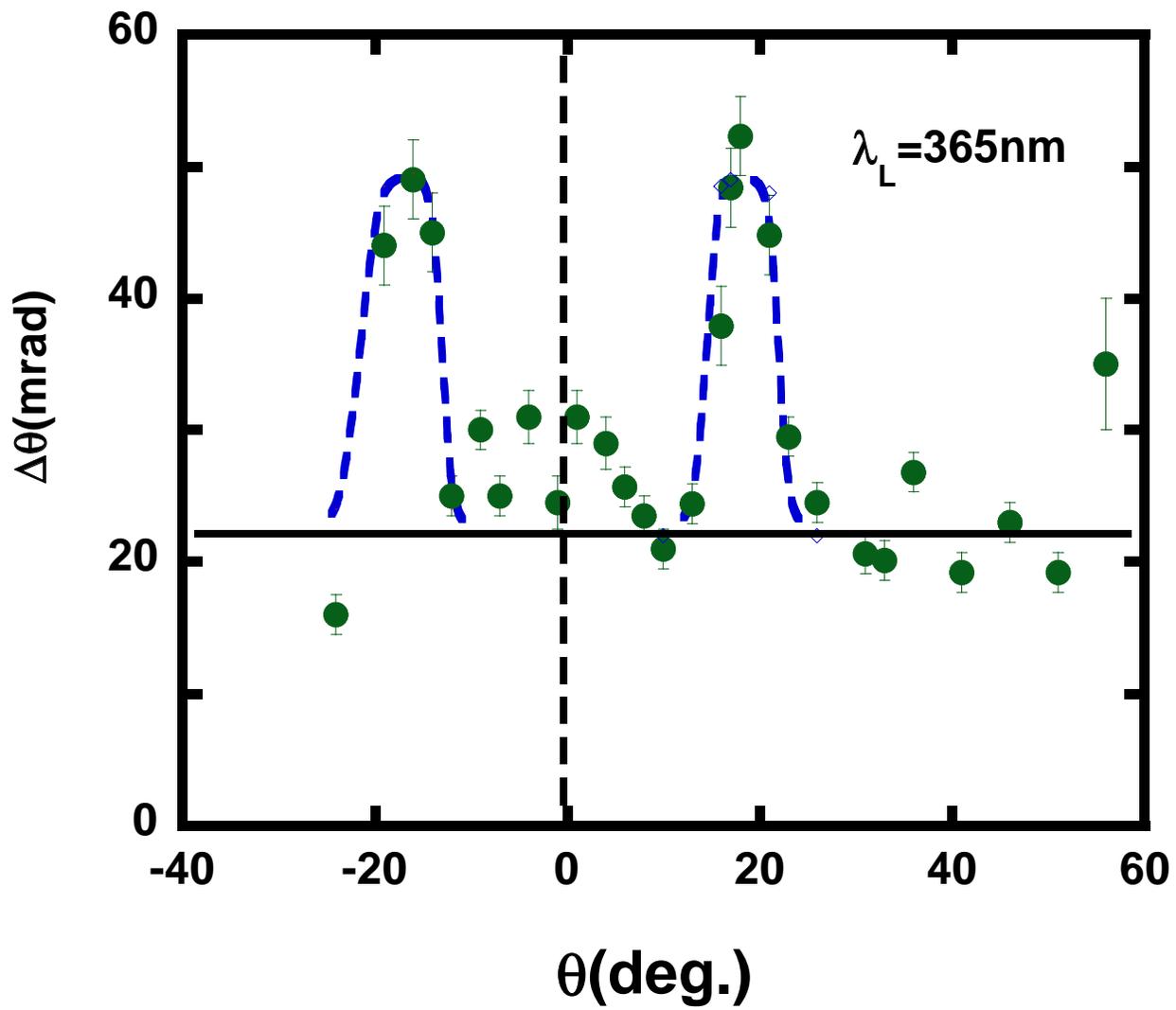
\end{document}